\documentclass[aps,twocolumn,superscriptaddress,longbibliography]{revtex4-1}

\usepackage{amsmath,amssymb,latexsym, url}
\usepackage{graphicx}
\usepackage{hyperref}
\usepackage{mathrsfs}
\usepackage{amscd}
\usepackage{color}
\renewcommand{\bm}[1]{\boldsymbol #1}

\usepackage{natbib}
\setcitestyle{authoryear,open={(},close={)}, sort}
\let\cite\citep

\begin{document}

\title{
Floquet States
}

\author{Naoto Tsuji}
\affiliation{Department of Physics, University of Tokyo, Hongo, Tokyo 113-0033, Japan}
\affiliation{RIKEN Center for Emergent Matter Science (CEMS), Wako 351-0198, Japan}

\begin{abstract}
Quantum systems driven by a time-periodic field are a platform of condensed matter physics 
where effective (quasi)stationary states, termed ``Floquet states'', can emerge with
external-field-dressed quasiparticles during driving. 
They appear, for example, as a prethermal intermediate state in isolated driven quantum systems
or as a nonequilibrium steady state in driven open quantum systems coupled to environment. 
Floquet states may have various intriguing physical properties, some of which can be drastically different 
from those of the original undriven systems in equilibrium. 
In this article, we review fundamental aspects of Floquet states, 
and discuss recent topics and applications of Floquet states in condensed matter physics.
\end{abstract}


\date{\today}

\maketitle


\section{Key objectives}

\begin{itemize}
\item Introduce Floquet states in periodically driven quantum systems.
\item Describe the basic aspects of the Floquet theory for isolated and open quantum systems.
\item Discuss several examples of Floquet states, including ac Wannier-Stark effects and Floquet topological phases.
\item Review experimental observations of Floquet states in solids and cold-atom systems.
\end{itemize}

\section{Introduction}

Driving quantum systems far from equilibrium provides various possibilities 
to control quantum states and realize new phases of matter that are otherwise inaccessible 
within thermal equilibrium. The interest on nonequilibrium states in condensed matter physics 
has grown rapidly due to recent progress both in theories and experiments, the latter of which allow for real-time observation 
of driven quantum states in a fast time scale in solids as well as in artificial quantum systems. 
There are many ways to drive quantum systems, 
among which time-periodic driving generates characteristic (quasi)stationary states called Floquet states,
which are often accompanied by quasiparticles dressed by external driving fields.
Their physical properties, such as band mass, lifetime, topological nature, etc., deviate from those of equilibrium states,
depending on the driving protocol.
This opens up a new avenue to control macroscopic phases dynamically.

The term, ``Floquet states'', originates from the Floquet theory for a set of ordinary differential equations
of the type, $\dot x_m(t)=\sum_n C_{mn}(t)x_n(t)$, with $C_{mn}(t)=C_{mn}(t+T)$ being a periodic function with a period $T$,
which dates back to the work of a mathematician, G. Floquet in 1883 \cite{Floquet1883, Hill1886, MagnusWinkler1966}.
According to the Floquet theory, there exists a solution in a form of $x_m(t)=e^{\lambda t} u_m(t)$ with a complex number $\lambda$
(characteristic exponent) and a periodic function $u_m(t)=u_m(t+T)$.
The application of the theory to the time-dependent Schr\"odinger equation 
in quantum mechanics laid down the foundation for the theory of Floquet states in periodically driven quantum systems 
\cite{Shirley1965, Sambe1973, Grifoni1998}. 
The Floquet theory can be viewed as a temporal analog of the Bloch theory in solid state physics,
which serves as a fundamental framework to treat electronic states in a spatially periodic crystal structure.

\begin{figure}[htbp]
\begin{center}
\includegraphics[width=7.5cm]{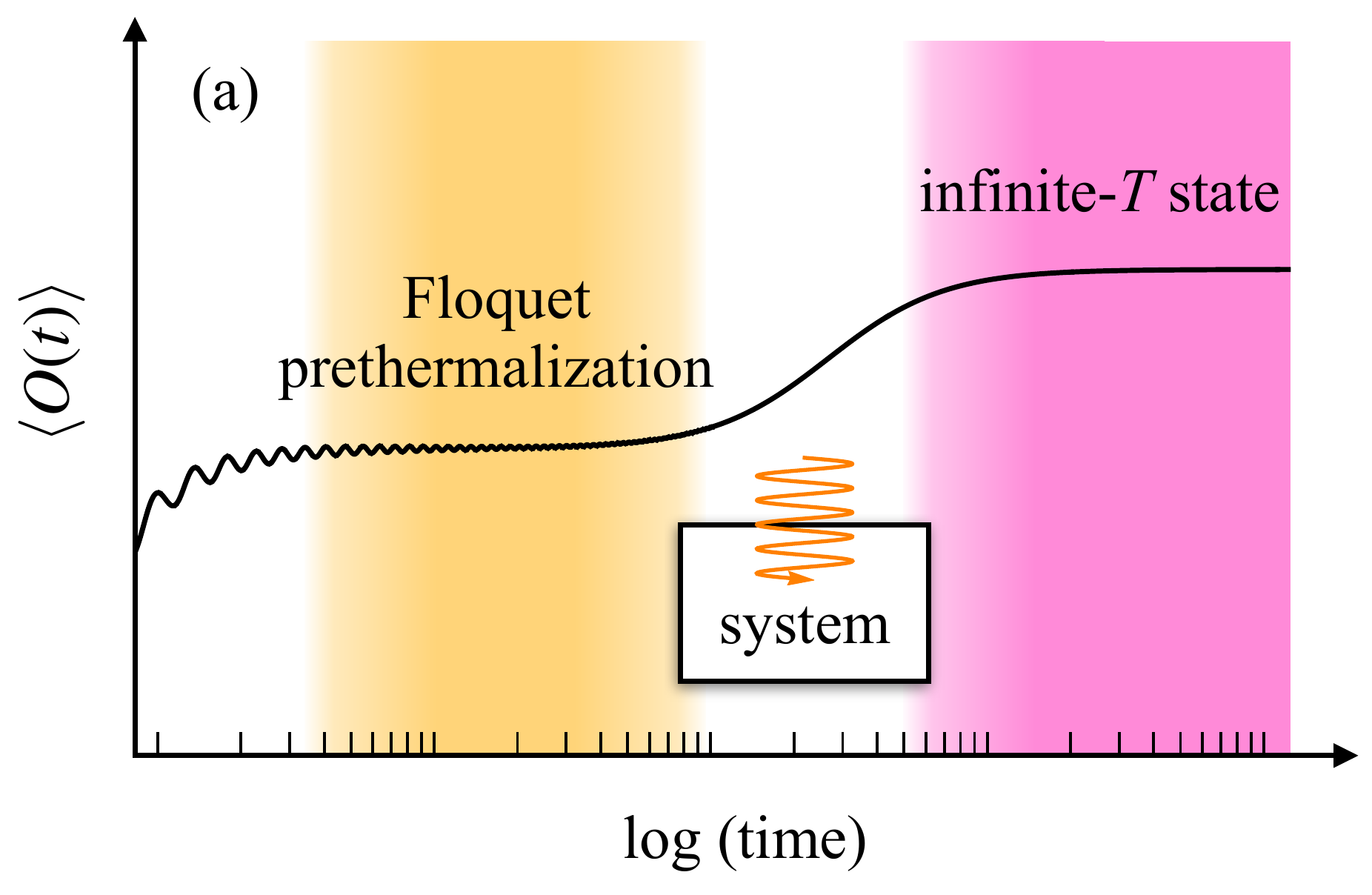}
\includegraphics[width=7.5cm]{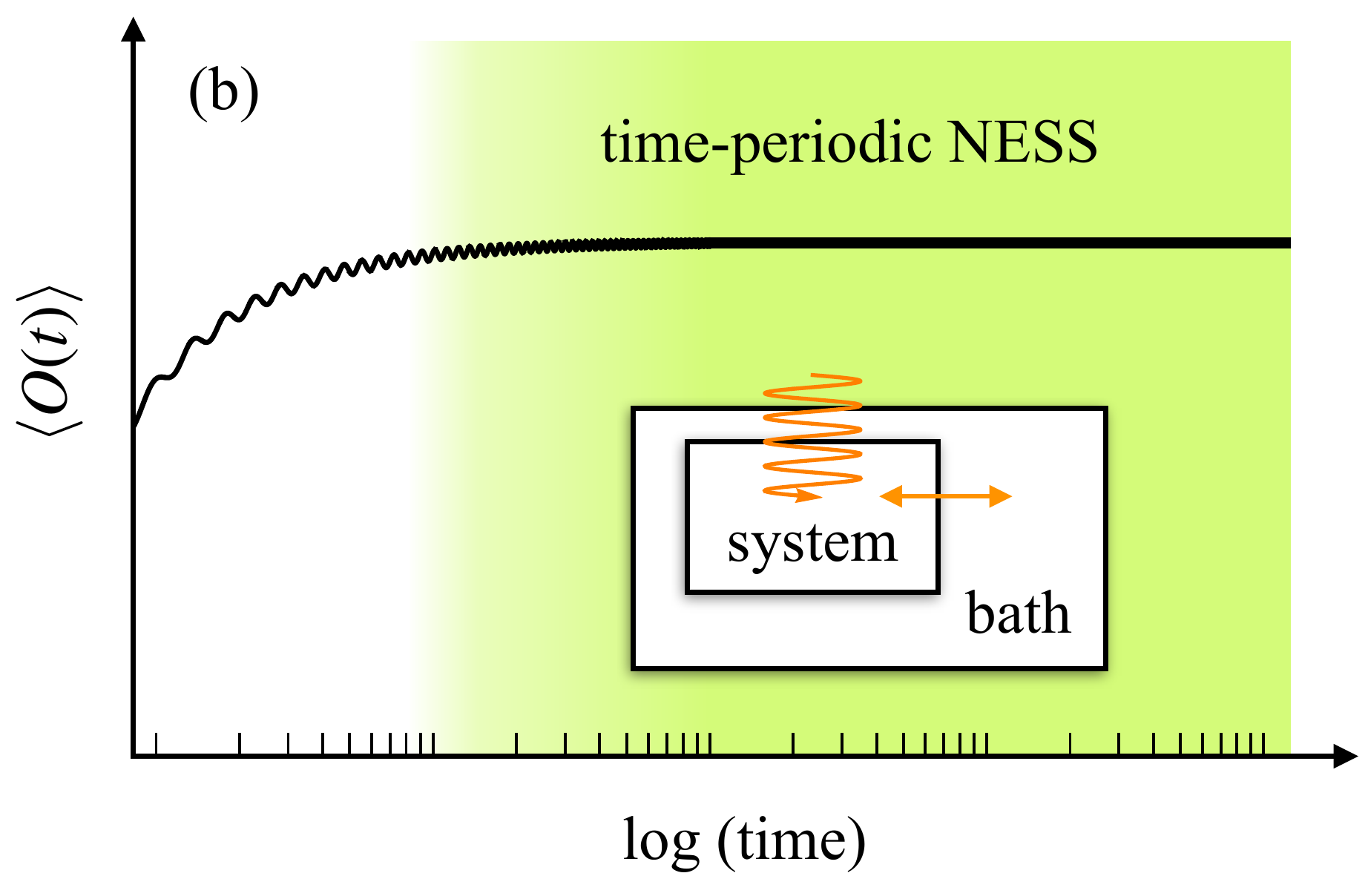}
\caption{Typical time evolution of an observable $\langle O(t)\rangle$
in (a) periodically driven isolated quantum systems and
(b) periodically driven open quantum systems. The former reaches an infinite-temperature steady state
through an intermediated quasistationary Floquet prethermal state, while the latter reaches
a time-periodic nonequilibrium steady state (NESS).}
\label{Floquet steady state}
\end{center}
\end{figure}

Periodically driven quantum systems are typically classified into two types: 
one is the case of isolated quantum systems driven by time-periodic fields, 
whose time evolution is purely determined by the unitary dynamics. 
The other is the case of open quantum systems, i.e., systems coupled to environment with dissipation. 
The dynamics are quite different between them (see Figure \ref{Floquet steady state}). 
The former eventually reaches an infinite-temperature state in the long-time limit \cite{Lazarides2014, DAlessioRigol2014, Ponte2015},
since the energy is continuously supplied to the system by the periodic driving, which turns into heat
in general (nonintegrable) systems.
Before approaching the infinite-temperature state, the system may stay at an intermediate quasi-stationary state 
(called Floquet prethermalization) for a certain time scale, 
which becomes exponentially long when the driving frequency is relatively high 
as compared to the system's energy scale \cite{Mori2016, Kuwahara2016, Abanin2017, Mori2018}. 
The latter, on the other hand, may arrive at 
a time-periodic nonequilibrium steady state (NESS), in which the energy absorbed from the external field is balanced 
with that dissipating into environment \cite{TsujiOkaAoki2009, OkaAoki2009, Dehghani2014, Mikami2016, Shirai2016, Murakami2017, IkedaSato2020}. 

In addition to these typical cases, nontrivial time-periodic stationary states
are known to emerge in driven many-body localized systems, where heating is
avoided without coupling to environment due to the effect of strong disorder \cite{Ponte2015b, Lazarides2015, Abanin2016}
and the emergence of approximately conserved charges \cite{Das2010, Haldar2018, Haldar2022}.
Recent studies \cite{Sacha2015, Khemani2016, Else2016, Yao2017, MoessnerSondhi2017}
have unveiled another possibility that time-periodic steady states 
in periodically driven quantum systems can possess a different periodicity than
that of the driving field, which is referred to as a (discrete) time crystal in the sense that
the discrete time translation symmetry is spontaneously broken (as in the space translation symmetry broken in a crystal).
In this way, the landscape of stationary states in periodically driven quantum systems is widely broadened,
and many different types of quasiparticles can be found there.

\section{Floquet theory}

In quantum mechanics, the Floquet theory plays a fundamental role
in describing Floquet states that emerge under an application of time-periodic driving fields.
In the case of isolated quantum systems, the time evolution is determined
by the time-dependent Schr\"odinger equation,
\begin{align}
i\frac{d}{dt}|\psi(t)\rangle
&=
H(t)|\psi(t)\rangle,
\label{Schrodinger equation}
\end{align}
where $|\psi(t)\rangle$ and $H(t)$ are the system's wavefunction and hermitian Hamiltonian, respectively,
and we set $\hbar=1$ throughout the article. 
The Hamiltonian is assumed to have time periodicity, $H(t+T)=H(t)$, with a period $T$
(and frequency $\omega=2\pi/T$).
The formal solution to the Schr\"odinger equation (\ref{Schrodinger equation}) is written as
$|\psi(t)\rangle=U(t,t_0)|\psi(t_0)\rangle$ ($t\ge t_0$),
where $U(t,t_0)$ is the unitary evolution operator 
defined by
\begin{align}
U(t,t_0)
&:=
\mathcal T\exp\left(-i\int_{t_0}^t d\bar t\, H(\bar t)\right)
\label{unitary evolution operator}
\end{align}
with $\mathcal T$ representing the time-ordered product.

The spirit of the Floquet theory lies in the idea of decomposing the overall dynamics into 
stroboscopic evolution (i.e., a sequence of snapshots at times $s\to s+T\to s+2T\to \cdots$) 
and micromotion (i.e., $s\to s+t$ ($0\le t<T$)).
The former contains information about long-time and time-averaged dynamics,
while the latter includes information on fast dynamics in a short-time scale.
To describe stroboscopic dynamics, it is convenient to consider time evolution by one cycle, 
starting from time $s$,
\begin{align}
U(s+T,s)
&=:
e^{-iH_F(s)T},
\label{H_F(s)}
\end{align}
where we define a hermitian operator $H_F(s)$ as an effective `static' Hamiltonian that describes time evolution over one period.
Alternatively, one can write $H_F(s)=iT^{-1}\log U(s+T,s)$. 
The eigenvalues of $H_F(s)$ are determined only up to modulo $\omega=2\pi/T$.
Once $H_F(s)$ is known, stroboscopic dynamics from $s$ to $s+nT$ ($n=1,2,3,\dots$) is 
given by $U(s+nT,s)=e^{-inH_F(s)T}$.
In general, however, $H_F(s)$ is a highly non-local and complicated many-body operator.


For general time evolution, one can rewrite the unitary operator $U(t,t_0)$ (\ref{unitary evolution operator}) as
\begin{align}
U(t,t_0)
&=
U(t,s)e^{i(t-s)H_F(s)}e^{-i(t-t_0)H_F(s)}
\notag
\\
&\quad\times
e^{-i(t_0-s)H_F(s)}U(s,t_0)
\end{align}
with a free parameter $s$. The above equation indicates how $U(t,t_0)$ differs from $e^{-i(t-t_0)H_F(s)}$.
If $t_0=s$ and $t=s+T$, the two are identical, while in other cases they are not in general. 
The discrepancy between $U(t,t_0)$ and $e^{-i(t-t_0)H_F(s)}$ is characterized by
a unitary operator
\begin{align}
P_s(t)
&:=
U(t,s)e^{i(t-s)H_F(s)},
\label{P_s(t)}
\end{align}
which is time periodic ($P_s(t+T)=P_s(t)$) by construction.
Using $P_s(t)$ (\ref{P_s(t)}), one can express the time-evolution operator as
\begin{align}
U(t,t_0)
&=
P_s(t) e^{-i(t-t_0)H_F(s)} P_s(t_0)^{-1}.
\end{align}

Now we are in a position to state Floquet's theorem in the present context.
\\

\noindent
{\bf (Floquet's theorem)}
{\it Given a time-periodic Hamiltonian $H(t)=H(t+T)$ and the corresponding unitary evolution operator 
$U(t,t_0)$} (\ref{unitary evolution operator}),
{\it there exist a time-independent hermitian operator $H_F$ and a time-periodic unitary operator $P(t)=P(t+T)$
such that
\begin{align}
U(t,t_0)
&=
P(t)e^{-i(t-t_0)H_F}P(t_0)^{-1}.
\label{Floquet's theorem}
\end{align}
}

Since we have already given an example of explicit constructions of the operators $H_F$ and $P(t)$ in the above
(Eqs.~(\ref{H_F(s)}) and (\ref{P_s(t)})),
the proof of the theorem has been completed.
The operator $H_F$ and its eigenvalues are often referred to as Floquet Hamiltonian
and quasienergies, respectively. 
If one substitutes Eq.~(\ref{Floquet's theorem}) in the equation of motion,
$i\partial_t U(t,t_0)=H(t)U(t,t_0)$, one obtains a relation between $H_F$ and $P(t)$,
\begin{align}
H_F
&=
P(t)^{-1}\left(H(t)-i\partial_t\right)P(t).
\label{H_F P(t)}
\end{align}

The choice of the combination $\{H_F, P(t)\}$ is not unique. This can be seen in the previous example, 
where $H_F(s)$ (\ref{H_F(s)}) and $P_s(t)$ (\ref{P_s(t)}) depend on the free parameter $s$.
On the other hand, $\{H_F, P(t)\}$ does not necessarily correspond to $\{H_F(s), P_s(t)\}$ for certain $s$.

Although the combination $\{H_F, P(t)\}$ is not unique, the eigenvalues of $H_F$ are unique (up to mod $\omega$).
Suppose that there exist two pairs, $\{H_F, P(t)\}$ and $\{\tilde H_F, \tilde P(t)\}$, that satisfy Eq.~(\ref{Floquet's theorem}).
By defining a unitary operator $V(t_0):=P(t_0)^{-1}\tilde P(t_0)$, one can see that 
they are related through
\begin{align}
e^{-i\tilde H_F T}
&=
V(t_0)^{-1}e^{-iH_F T}V(t_0),
\\
\tilde P(t)
&=
P(t)e^{-i(t-t_0)H_F}V(t_0)e^{i(t-t_0)\tilde H_F}.
\label{tilde P}
\end{align}
These are a kind of `gauge transformations' that characterize non-uniqueness of $\{H_F, P(t)\}$.
Note that $e^{-i\tilde H_FT}$ is related to $e^{-iH_FT}$ through the unitary transformation,
hence the eigenvalues of $H_F$ and $\tilde H_F$ are identical up to mod $\omega$
(except for the difference in order).
It is always possible to choose a gauge in such a way that $P(t_0)=I$ (the identity operator). 
If $P(t_0)\neq I$, one can perform a gauge transformation
with $V(t_0)=P(t_0)^{-1}$, which gives $\tilde P(t_0)=I$.
In this particular gauge, the unitary operator is simplified to $U(t,t_0)=P(t)e^{-i(t-t_0)H_F}$.

The wavefunction evolves, according to Floquet's theorem (\ref{Floquet's theorem}), as
$|\psi(t)\rangle=P(t)e^{-i(t-t_0)H_F}P(t_0)^{-1}|\psi(t_0)\rangle$.
If $P(t_0)^{-1}|\psi(t_0)\rangle$ is an eigenstate (say $|\psi_\alpha\rangle$) of $H_F$ (with an eigenvalue $\varepsilon_\alpha$),
the wavefunction takes a form of
\begin{align}
|\psi(t)\rangle
&=
e^{-i\varepsilon_\alpha(t-t_0)}|u_\alpha(t)\rangle
\label{Floquet wavefunction}
\end{align}
with
$|u_\alpha(t)\rangle:=P(t)|\psi_\alpha\rangle$,
which is time periodic, $|u_\alpha(t+T)\rangle=|u_\alpha(t)\rangle$.
Thus, the wavefunction is decomposed into the phase factor and the time-periodic part in Eq.~(\ref{Floquet wavefunction}).
This is analogous to the well-known Bloch's theorem for spatially periodic systems in solid-state physics, in which
the Bloch wavefunction has a form of the product of a plane wave and a spatially periodic function
(see Table~\ref{Bloch Floquet} for comparison).

If $P(t_0)^{-1}|\psi(t_0)\rangle$ is not an eigenstate of $H_F$, one can still expand it 
in the complete eigenbasis of the hermitian operator $H_F$, $\{ |\psi_\alpha\rangle \}_\alpha$ , as
\begin{align}
P(t_0)^{-1}|\psi(t_0)\rangle
&=
\sum_\alpha c_\alpha |\psi_\alpha\rangle.
\end{align}
Then the wavefunction at time $t$ is given by
\begin{align}
|\psi(t)\rangle
&=
\sum_\alpha c_\alpha e^{-i\varepsilon_\alpha (t-t_0)}|u_\alpha(t)\rangle,
\end{align}
where $c_\alpha$ is a time-independent coefficient.
Again, the wavefunction is expressed as a combination of the phase factors and the time-periodic functions
$|u_\alpha(t)\rangle$. 

\begin{table}[t]
\begin{tabular}{ccc}
\hline
& Bloch & Floquet \\
\hline
\hline
Hamiltonian & $H(\bm r+\bm R)=H(\bm r)$ & $H(t+T)=H(t)$ \\
wavefunction & $|\psi_{\bm k}(\bm r)\rangle=e^{i\bm k\cdot\bm r}|u_{\bm k}(\bm r)\rangle,$
& $|\psi_\alpha(t)\rangle=e^{-i\varepsilon_\alpha t}|u_\alpha(t)\rangle,$ \\
& $|u_{\bm k}(\bm r+\bm R)\rangle=|u_{\bm k}(\bm r)\rangle$ & $|u_\alpha(t+T)\rangle=|u_\alpha(t)\rangle$ \\
& (crystal momentum $\bm k$) & (quasienergy $\varepsilon_\alpha$) \vspace{.1cm} \\
Brillouin zone & $\displaystyle -\frac{\pi}{a}\le \bm k_i< \frac{\pi}{a}$ & $\displaystyle -\frac{\omega}{2}\le\varepsilon<\frac{\omega}{2}$
\vspace{.1cm} \\
\hline
\end{tabular}
\caption{Comparison between Bloch's theorem for spatially periodic systems and Floquet's theorem for time-periodic systems.
The Brillouin zone in momentum space is shown for a simple cubic lattice with a lattice constant $a$ as an example.}
\label{Bloch Floquet}
\end{table}

An observable $\langle O(t)\rangle=\langle\psi(t)| O |\psi(t)\rangle$ is not necessarily time periodic
due to the presence of off-diagonal terms in the Floquet eigenbasis, 
$e^{i(\varepsilon_\alpha-\varepsilon_\beta)(t-t_0)}\langle u_\alpha(t)|O|u_\beta(t)\rangle$ ($\alpha\neq\beta$). 
However, there are various situations where the contributions of the off-diagonal terms are suppressed,
such as Floquet prethermalized states in an isolated system and nonequilibrium steady states
in open systems (Figure \ref{Floquet steady state}). In those cases, 
time evolution of observables becomes periodic with the same periodicity as the driving field.
An important exception is a discrete time crystal
\cite{Sacha2015, Khemani2016, Else2016, Yao2017, MoessnerSondhi2017}, 
where observables become time periodic
but its periodicity is integer multiples of that of the driving field.

One can look at Floquet's theorem from another point of view, which is even closer to that of solid-state physics,
i.e., the band picture. 
By substituting the Floquet wavefunction (\ref{Floquet wavefunction}) into the time-dependent
Schr\"odinger equation (\ref{Schrodinger equation}), one obtains
$i\frac{d}{dt}|u_\alpha(t)\rangle+\varepsilon_\alpha|u_\alpha(t)\rangle=H(t)|u_\alpha(t)\rangle$.
Since $|u_\alpha(t)\rangle$ and $H(t)$ are both periodic in time, one can Fourier transform them
into Fourier modes,
\begin{align}
H_{mn}
&:=
T^{-1}\int_0^T dt\, e^{i(m-n)\omega t}H(t),
\label{H_mn}
\\
|u_\alpha^n\rangle
&:=
T^{-1}\int_0^T dt\, e^{in\omega t}|u_\alpha(t)\rangle.
\label{u_alpha^n}
\end{align}
Since $H_{mn}$ only depends on $m-n$, one can also use a shorthand notation of $H_{m-n}:=H_{mn}$.
After that, the time-dependent Schr\"odinger equation turns into a seemingly time-independent equation,
\begin{align}
&
\sum_n \mathcal (\mathcal H_F)_{mn}|u_\alpha^n\rangle
=
\varepsilon_\alpha|u_\alpha^m\rangle,
\label{Floquet eigenvalue equation}
\\
&
(\mathcal H_F)_{mn}
:=
H_{mn}-m\omega\delta_{mn}.
\label{H_F}
\end{align}


Here comes an idea of extending the Hilbert space in order to interpret Eq.~(\ref{Floquet eigenvalue equation})
as an eigenvalue problem.
Let $\mathbb H$ be the Hilbert space of the system, and let $\mathbb T$ be a vector space
spanned by Fourier modes of time-periodic functions with the period $T$. For example, 
a periodic function $f(t)=f(t+T)$ is Fourier transformed as $f(t)=\sum_n f_n e^{-in\omega t}$.
Then a vector $(\cdots, f_{-2}, f_{-1}, f_0, f_1, f_2, \cdots)$, whose elements are composed of Fourier coefficients,
belongs to $\mathbb T$. Equation (\ref{Floquet eigenvalue equation}) can be viewed as an eigenvalue equation
in the extended Hilbert space $\mathbb H \otimes \mathbb T$ (Sambe space \cite{Sambe1973}).

The set $\{\varepsilon_\alpha\}_\alpha$ has been defined as the eigenvalues of $H_F$ acting on $\mathbb H$, 
while in Eq.~(\ref{Floquet eigenvalue equation})
they appear as the eigenvalues of $\mathcal H_F$ acting on $\mathbb H\otimes\mathbb T$.
Since the dimensions of $\mathbb H$ and $\mathbb H\otimes\mathbb T$ are different, 
there must be missing eigenvalues of $\mathcal H_F$,
which can be found if one notices the fact that a vector $|u_\alpha^{n-l}\rangle$ shifted from $|u_\alpha^n\rangle$ 
by $l$ components ($l=0,\pm 1,\pm 2,\dots$)
is also an eigenstate of $\mathcal H_F$ with an eigenvalue $\varepsilon_\alpha-l\omega$.
This means that $\{|u_\alpha^{n-l}\rangle \}_{\alpha, l}$ consists of a complete set of the eigenstates 
in $\mathbb H\otimes\mathbb T$. The spectrum of the corresponding eigenvalues $\{\varepsilon_\alpha-l\omega\}_{\alpha, l}$ 
shows a periodic structure in energy space, in much the same way as the band structure shows a periodic pattern
in momentum space for spatially periodic systems. In particular, one can restrict the energy space 
within $-\frac{\omega}{2}\le \varepsilon_\alpha <\frac{\omega}{2}$
(the Floquet Brillouin zone) by shifting quasienergies by integer multiples of $\omega$
in order to represent quasienergies without mod $\omega$ redundancies.

Thus, $H_F$ and $\mathcal H_F$ share the same eigenvalues up to mod $\omega$.
As we have mentioned, $H_F$ is a complicated non-local many-body operator that is expected to be difficult to evaluate directly,
while $\mathcal H_F$ is usually a well-behaved local few-body operator. On the other hand,
the latter is acting on the extended space $\mathbb H\otimes\mathbb T$, whose dimension is infinitely larger than $\mathbb H$.
In practice, one can truncate the Fourier space $\mathbb T$ to finite dimensions as an approximation;
Off-diagonal elements, $(\mathcal H_F)_{mn}$, physically correspond to $m-n$ photon absorption/emission processes,
which would be suppressed when $|m-n|\gg 1$ if the drive strength is not too strong.

\begin{figure}[t]
\includegraphics[width=7cm]{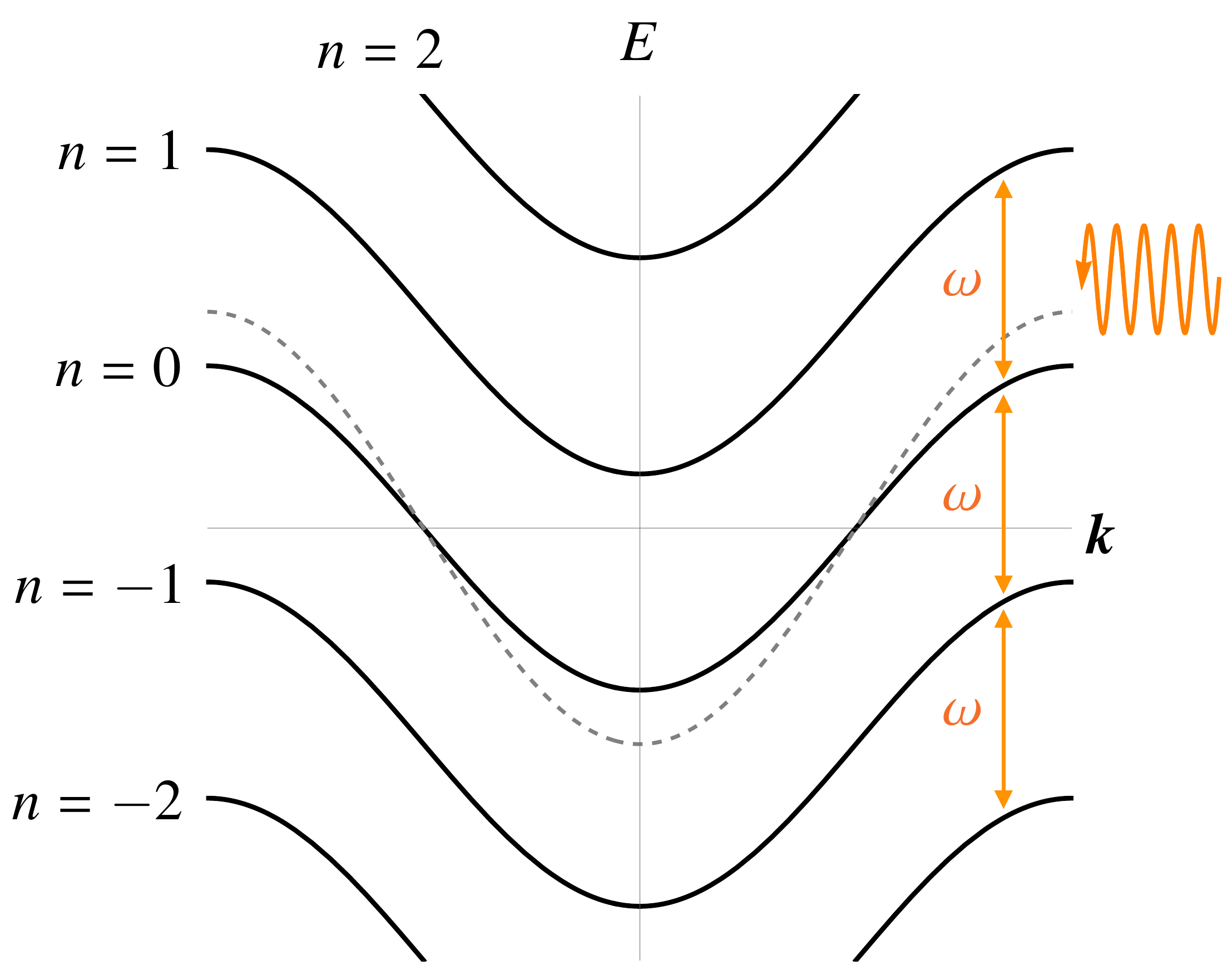}
\caption{The quasienergy spectrum of noninteracting electrons on a one-dimensional lattice
driven by an ac electric field (solid curves). The dashed curve shows the original energy dispersion without the driving field.}
\label{ac Wannier Stark}
\end{figure}

\section{Examples}
\label{sec: examples}

As an example, let us consider a single-band system of noninteracting electrons
driven by a time-periodic field described by the Hamiltonian
\begin{align}
H(t)
&=
\sum_{\bm k} \varepsilon_{\bm k}(t)c_{\bm k}^\dagger c_{\bm k},
\label{single-band Hamiltonian}
\end{align}
where $c_{\bm k}^\dagger$ is a creation operator of electrons with momentum $\bm k$,
and $\varepsilon_{\bm k}(t)=\varepsilon_{\bm k}(t+T)$ is a band dispersion including the coupling to a time-periodic field.
The eigenvalues (quasienergies) of the corresponding $\mathcal H_F$ (\ref{H_F}) are given by
$T^{-1}\int_0^T dt\, \varepsilon_{\bm k}(t)+n\omega$ ($n=0,\pm1,\pm2,\dots$)
\cite{TsujiOkaAoki2008},
i.e., the time-averaged dispersion shifted by integer multiples of the drive frequency $\omega$.
One can confirm that these are also eigenvalues of $H_F$ (\ref{Floquet's theorem}),
since the time evolution operator is given by 
$U(t,t_0)=e^{-i\int_{t_0}^t ds\, \sum_{\bm k}\varepsilon_{\bm k}(s)c_{\bm k}^\dagger c_{\bm k}}$
which is free from the time-ordering operator due to $[H(t), H(t')]=0$.

If the system is a one-dimensional lattice with a nearest-neighbor hopping with a dispersion $\varepsilon_k=-2J\cos k$
driven by an ac electric field with amplitude $E$ (represented by a vector potential $A(t)=-E\sin\omega t/\omega$), 
the time average of the dispersion is given by $T^{-1}\int_0^T dt\, (-2J)\cos(k-A(t))=-2J\mathcal J_0(E/\omega)\cos k$,
where the hopping amplitude $J$ is multiplied by $\mathcal J_0(E/\omega)$, the zeroth-order Bessel-function factor.
In Figure~\ref{ac Wannier Stark}, we show the corresponding quasienergy spectrum.
Due to the presence of the periodic drive, the hopping amplitude is effectively reduced ($|\mathcal J_0(E/\omega)|\le 1$),
and electrons become heavier. When the Bessel factor exactly vanishes ($\mathcal J_0(E/\omega)=0$), 
electrons cannot hop to neighboring sites completely, resulting in dynamical localization \cite{Dunlap1986, Holthaus1992}. 
In the spectrum (Fig.~\ref{ac Wannier Stark}), there are side-band structures (labeled by the Fourier mode index $n$) separated by $\omega$
from the neighboring bands, which is referred to as the dynamic Wannier-Stark ladder. 
These arise due to the consequence of virtual $n$-photon absorption/emission processes.

\begin{figure}[t]
\includegraphics[width=7.5cm]{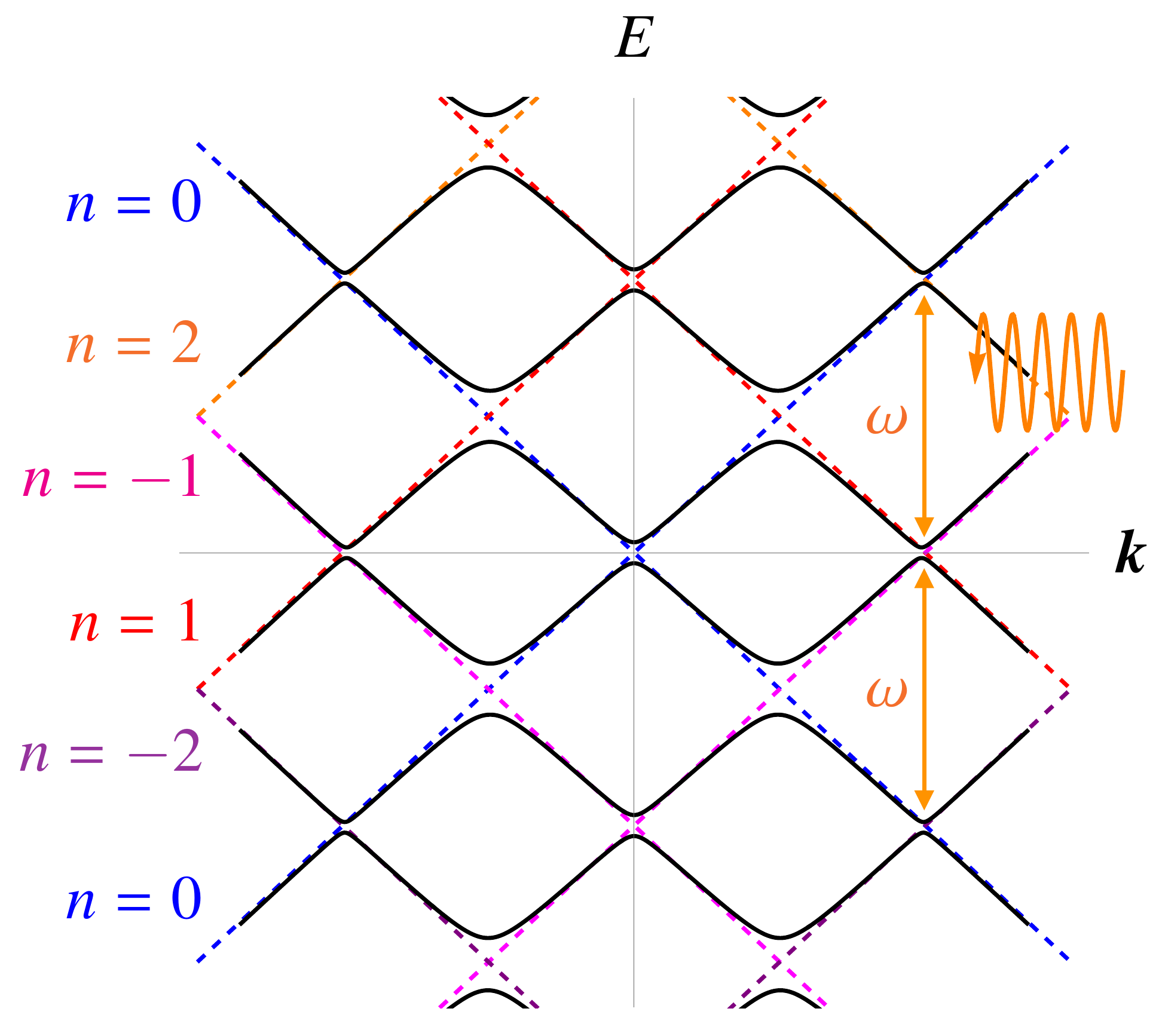}
\caption{The quasienergy spectrum (solid curves) of Dirac electrons driven by circularly polarized light
($A=1, \omega=5$).
The dashed lines with Fourier mode labels $n$ show the spectrum without driving.}
\label{dirac}
\end{figure}
Another example is a model of Dirac electrons in two spatial dimensions driven by circularly polarized light
\cite{OkaAoki2009, Mikami2016}.
The Hamiltonian reads
\begin{align}
H(t)
&=
\sum_{\bm k\alpha\beta} c_{\bm k\alpha}^\dagger h_{\bm k\alpha\beta}(t) c_{\bm k\beta},
\\
h_{\bm k\alpha\beta}(t)
&=
(k_x-A_x(t))(\sigma_x)_{\alpha\beta}+(k_y-A_y(t))(\sigma_y)_{\alpha\beta},
\end{align}
where $c_{\bm k\alpha}^\dagger$ is a creation operator of electrons with two components
(corresponding to two bands near the Fermi energy),
$(A_x, A_y)=(A\cos\omega t,A\sin\omega t)$ is the vector potential for circularly polarized light, 
and $\sigma_x$ and $\sigma_y$ are Pauli matrices.
In this system, circularly polarized light explicitly breaks time reversal symmetry.

In Figure~\ref{dirac}, a typical quasienergy spectrum is shown.
When the driving field is absent, the spectrum consists of a Dirac cone (linear dispersion) without energy gap near $E=0$.
As one increases the amplitude of the electric field, there emerge energy gaps at the band crossing points, accompanied by
sideband structures with energy spacing $\omega$.
Especially, at $\bm k=0$ the gap is given by $\Delta=\sqrt{\omega^2+4A^2}-\omega$ \cite{OkaAoki2009}.
This phenomenon, light-induced gap opening in the quasienergy spectrum (accompanied by chiral edge modes), 
can be interpreted as a topological phase transition,
since quasienergy bands are characterized by non-zero topological (Chern) numbers,
similar to quantum Hall states in static systems.
In this sense, the system is classified to a Floquet topological insulator \cite{OkaAoki2009, Kitagawa2010, Kitagawa2011, Lindner2011, Roy2017, Rudner2020}, 
a class of topological phases in periodically driven systems.

While the above example can be understood by mapping to the corresponding static model, 
there are several examples that show nontrivial topological phases that do not have equilibrium counterparts
\cite{Kitagawa2010, Kitagawa2012, Rudner2013, Harper2020, Nakagawa2020b, Jangjan2022}:
Topologically protected chiral edge modes can appear even though all the bulk Floquet bands have vanishing Chern numbers.
This is possible since in Floquet systems edge modes can go across the quasienergy $\varepsilon=\pm\pi/T$ (`anomalous $\pi$ modes')
and wind around the Floquet Brillouin zone in the frequency domain. 
In this case, the system is called an anomalous Floquet topological insulator.


\section{High-frequency expansion}
\label{sec: high-frequency expansion}

The approach of the Floquet theory becomes particularly effective when the drive frequency is higher than
typical energy scales included in the original (un-driven) Hamiltonian. Intuitively, when the frequency is sufficiently large,
physical degrees of freedom in the system cannot follow rapid oscillations of the drive field.
As a result, the effect of the drive is rounded in the time average, implying that
the Floquet Hamiltonian $H_F$ defined in Floquet's theorem (\ref{Floquet's theorem}) 
is approximated to be the time average of the original Hamiltonian, $H_F\approx H_0=T^{-1}\int_0^T dt\, H(t)$.
This intuition is actually justified in the form of high-frequency expansions (or $1/\omega$ expansion),
a typical one of which reads
\begin{align}
H_F
&=
H_0+\sum_{n=1}^\infty \frac{[H_{-n},H_n]}{n\omega}+\cdots,
\label{high-frequency expansion}
\end{align}
where ``$\cdots$'' contains higher-order terms in $1/\omega$. The second term on the right-hand side of Eq.~(\ref{high-frequency expansion})
can be understood as a sum of second-order perturbation corrections for $n$-photon absorption/emission processes. 

There are several ways to carry out such an expansion
(e.g., Floquet-Magnus expansion \cite{Feldman1984, Casas2001, Blanes2009, Mananga2011, Goldman2014}, 
van Vleck perturbation theory \cite{Eckardt2015, Bukov2015}, 
Brillouin-Wigner perturbation theory \cite{Mikami2016}, etc.), 
all of which are known to be equivalent to each other in the sense 
that they all give the same high-frequency expansion for the eigenvalues of $H_F$ (up to mod $\omega$).
One can derive some of them, for example, by starting from Eq.~(\ref{H_F P(t)}),
$H_F=
e^{i\Lambda(t)}(H(t)-i\partial_t)e^{-i\Lambda(t)}$,
where $\Lambda(t)$ is a hermitian operator defined by $P(t)=: e^{-i\Lambda(t)}$.
This relation can be rewritten with an adjoint operator ${\rm ad}_{\Lambda(t)} X:=[\Lambda(t), X]$ as
\begin{align}
\frac{d}{dt}\Lambda(t)
&=
\sum_{n=0}^\infty \frac{B_n}{n!}(-i)^n ({\rm ad}_{\Lambda(t)})^n [H(t)+(-1)^{n+1}H_F],
\label{d/dt Lambda(t)}
\end{align}
where $B_n$ is the Bernoulli number.
One can solve Eq.~(\ref{d/dt Lambda(t)}) order-by-order by expanding
\begin{align}
\Lambda(t)
&=
\sum_{k=1}^\infty \Lambda^{(k)}(t),
\quad
H_F
=
\sum_{k=0}^\infty H_F^{(k)},
\end{align}
where one assigns an order 1 for $H(t)$, the order $k$ for $\Lambda^{(k)}$, and the order $k+1$ for $H^{(k)}$.
In this rule, the superscript $(k)$ of $\Lambda^{(k)}$ and $H^{(k)}$ represents the order in $1/\omega$.

To determine the expansion, one needs to fix the gauge.
For example, one can set $\Lambda(t_0)=0$ at certain time $t_0$, which generates the so-called Floquet-Magnus expansion.
While the expansion can be uniquely determined, there remains a spurious dependence on $t_0$ 
that appears as a phase factor in the expansion.
Another choice that is often employed is $\int_0^T dt\, \Lambda(t)=0$,
which corresponds to the van Vleck perturbation expansion (Eq.~\ref{high-frequency expansion})
that does not have explicit $t_0$ dependence.
These two expansions are related through unitary transformation.

The high-frequency expansions are usually asymptotic expansions. That is,
they do not converge in the ordinary sense but give good approximation if the infinite series are truncated
at a certain finite order. In fact, there is a rigorous estimate for the error induced by such a truncation,
which suggests that the truncated Floquet-Magnus expansion accurately describes time evolution
of periodically driven quantum systems up to a time scale which is exponentially long
as a function of drive frequency. As a result, there emerges an exponentially long-lived prethermal regime
(Fig.~\ref{Floquet steady state}(a)) at an early stage of periodic driving,
where the system shows thermal behavior with respect to the effective Hamiltonian given by
the truncated expansion.

As an example, let us apply the high-frequency expansion to a model of noninteracting electrons on a honeycomb lattice
driven by circularly polarized light. The Hamiltonian is given by
\begin{align}
H(t)
&=
J\sum_{\langle jk\rangle} 
\left[\exp\left(-i\int_{\bm R_k}^{\bm R_j} \bm A(t)\cdot d\bm r\right) c_j^\dagger c_k+\mbox{h.c.}\right],
\end{align}
where $\langle jk\rangle$ denotes nearest-neighbor sites at positions $\bm R_j$ and $\bm R_k$
on the honeycomb lattice, and $\bm A(t)$ represents
a vector potential for circularly polarized light, which is included as a Peierls phase.
In the high-frequency expansion (\ref{high-frequency expansion}), the corresponding effective Hamiltonian is given by
\cite{Mikami2016}
\begin{align}
H_F
&=
J_{\rm eff}\sum_{\langle jk\rangle} (c_j^\dagger c_k+\mbox{h.c.})
+K_{\rm eff}\sum_{\langle\!\langle jk\rangle\!\rangle} (i\tau_{j,k}c_j^\dagger c_k+\mbox{h.c.})
\notag
\\
&\quad
+O\left(\frac{1}{\omega^2}\right),
\end{align}
where $\langle\!\langle jk\rangle\!\rangle$ represents next-nearest-neighbor sites,
$J_{\rm eff}=J \mathcal J_0(A)$, $K_{\rm eff}=-J^2/\omega \sum_{n\neq 0} \mathcal J_n(A)^2 \sin(2\pi n/3)/n$
($\mathcal J_n(A)$ is the $n$th-order Bessel function), and $\tau_{j,k}=\pm 1$ represents
the chirality of the hopping from $k$ to $j$ ($+$($-$) is assigned to the clockwise (counterclockwise) path on a hexagon).

In addition to the modulated nearest-neighbor hopping, there appear next-nearest-neighbor hoppings that are purely imaginary.
The resulting Hamiltonian is equivalent to the Haldane model \cite{Haldane1988}, a tight-binding model on a honeycomb lattice
with complex next-nearest-neighbor hoppings, which is known to be a model of quantum anomalous Hall states (Chern insulators).
Thus, one can understand that the system undergoes a topological phase transition induced by circularly polarized light.
This is consistent with the result of the model of Dirac electrons treated in the previous section.
As one reduces the drive frequency, higher-order terms in the high-frequency expansion become relevant, 
and more complicated topological phases can be found with high Chern numbers \cite{Mikami2016}.

\section{Periodically driven open quantum systems}

In open quantum systems, the time evolution of the system is not simply given 
by the time-dependent Schr\"odinger equation (\ref{Schrodinger equation}) due to the effect of environment.
Let the total Hamiltonian be given as $H=H_S+H_I+H_B$, where $H_S$ is the system's Hamiltonian,
$H_I$ represents the interaction between the system and environment,
and $H_B$ is the Hamiltonian of environment (bath). The density matrix of the total system $\rho$ evolves
according to the von Neumann equation, $\frac{d}{dt}\rho=-i[H, \rho]$.
What one is interested in here is the time evolution of the system's density matrix defined by $\rho_S={\rm tr}_B \rho$
(${\rm tr}_B$ denotes the partial trace over environment's degrees of freedom).

One approach to describe the system's dynamics is to use a quantum Master equation, which is a quantum analog
of Master equations in classical stochastic processes \cite{BreuerPetruccione}. To derive the quantum Master equation,
there are two basic assumptions: (i) Markovianity of system's time evolution
(i.e., $\rho_S(t+\Delta t)$ is solely determined from $\rho_S(t)$), 
and (ii) absence of initial correlations between
the system and environment (i.e., $\rho(0)=\rho_S(0)\otimes\rho_B(0)$).
In this situation, if a map from arbitrary $\rho_S(0)$ to $\rho_S(t)$ is completely positive and trace preserving,
$\rho_S(t)$ obeys the following form of the quantum Master equation
(Gorini-Kossakowski-Sudarshan-Lindblad equation \cite{GKS1976, Lindblad1976}),
\begin{align}
\frac{d}{dt}\rho_S
&=
-i[H_S, \rho_S]+\sum_j \left(
L_j \rho_S L_j^\dagger-\frac{1}{2}\{L_j^\dagger L_j, \rho_S\}
\right),
\label{Lindblad equation}
\end{align}
with a certain set of operators $\{L_j\}$ called Lindblad operators (or jump operators). The second term
on the right-hand side of Eq.~(\ref{Lindblad equation}) represents the effect of environment,
giving non-unitary dynamics. If one defined a non-hermitian effective Hamiltonian,
$H_{\rm eff}:=H_S-\frac{i}{2}\sum_j L_j^\dagger L_j$, Eq.~(\ref{Lindblad equation}) is rewritten as
\begin{align}
\frac{d}{dt}\rho_S
&=
-i(H_{\rm eff}\rho_S-\rho_S H_{\rm eff}^\dagger)+\sum_j L_j \rho_S L_j^\dagger.
\label{Lindblad non-hermitian}
\end{align}
The second term on the right-hand side of Eq.~(\ref{Lindblad non-hermitian}) represents quantum jump processes,
in which the system's state discontinuously changes in a stochastic manner. 

A microscopic justification of the quantum Master equation (\ref{Lindblad equation}) often requires additional approximations. 
The conventional approach employs (i) the Born approximation which is valid for the weak coupling between the system and environment
and (ii) the secular approximation (or the rotating-wave approximation) that can be justified when the inverse of the level spacing between the eigenvalues of $H_S$
is much smaller than the relaxation time of the system. The latter condition does not usually hold for many-body systems, since the level spacing
becomes exponentially small as the system size increases.
Nevertheless, the quantum Master equation has been used from a phenomenological point of view to describe the dynamics of open quantum many-body systems \cite{Diehl2010, Prosen2011, Tindall2019, Nakagawa2020, Yamamoto2021}.

In the presence of periodic driving, the use of the quantum Master equation needs more careful derivation \cite{Mori2023}.
Especially, if one is interested in nontrivial nonequilibrium steady states, heating due to the periodic drive should be balanced with dissipation.
The relaxation time then becomes comparable to the system's intrinsic time scale, which may invalidate the secular approximation.
There has recently been a proposal for the derivation of quantum Master equations without relying on the secular approximation 
(or the rotating-wave approximation) \cite{Nathan2020, Nathan2022}. The underlying notion is that Markovianity is a property of 
the bath and the system-bath coupling solely, and should not depend on details of the energy level structure of the system. 
In fact, one can derive a Markovian quantum Master equation (called the `universal Lindblad equation')
when the correlation time of the bath is much shorter than the characteristic time scale of the system-bath interaction.

Another approach to driven open quantum systems is to use nonequilibrium Green's functions \cite{KadanoffBaymBook, Keldysh1964}.
The advantage of this approach is that it does not necessarily rely on the Markov and secular approximations.
On the other hand, since the approach is based on the Green's function formalism, one has to adopt certain diagrammatic techniques
to treat many-body interactions.

One of the simplest models of environment for electrons is a free-fermion heat bath \cite{TsujiOkaAoki2009, Aoki2014},
whose Hamiltonians are given by
\begin{align}
H_B
&=
\sum_{j,k} \varepsilon_k b_{j,k}^\dagger b_{j,k},
\\
H_I
&=
\sum_{j,k} V_k (c_j^\dagger b_{j,k}+{\rm h.c.}),
\end{align}
where $c_j^\dagger$ and $b_{j,k}^\dagger$ are creation operators of system's and bath's electrons at site $j$ and mode $k$, respectively,
$\varepsilon_k$ is the energy of bath's electrons, and $V_k$ is a coupling between system's
and bath's electrons. 
The system's retarded, advanced, and Keldysh Green's functions are defined as
\begin{align}
G_{ij}^R(t,t')
&=
-i\theta(t-t')\langle \{c_i(t), c_j^\dagger(t')\}\rangle,
\\
G_{ij}^A(t,t')
&=
i\theta(t'-t)\langle \{c_i(t), c_j^\dagger(t')\}\rangle,
\\
G_{ij}^K(t,t')
&=
-i\langle [c_i(t), c_j^\dagger(t')]\rangle,
\end{align}
respectively ($\theta(t)$ is the step function). The retarded (and advanced) Green's function contains information about
the single-particle spectrum of the system, while the Keldysh one carries information about the occupation distribution of the spectrum.

Since the total Hamiltonian $H=H_S+H_I+H_B$ is quadratic in $b_{j,k}^\dagger$ and $b_{j,k}$,
one can integrate out the bath's degrees of freedom analytically. As a result, the system's Green's function
satisfies the following Dyson equation,
\begin{align}
\begin{pmatrix}
G^R & G^K \\
0 & G^A
\end{pmatrix}
&=
\Bigg[
\begin{pmatrix}
G_0^R & G_0^K \\
0 & G_0^A
\end{pmatrix}^{-1}
-
\begin{pmatrix}
\Sigma_b^R & \Sigma_b^K \\
0 & \Sigma_b^A
\end{pmatrix}
\notag
\\
&\quad
-
\begin{pmatrix}
\Sigma^R & \Sigma^K \\
0 & \Sigma^A
\end{pmatrix}
\Bigg]^{-1},
\label{Dyson equation}
\end{align}
where $G_0$ is the system's noninteracting Green's function,
$\Sigma_b$ is the self-energy due to the coupling to the heat bath,
and $\Sigma$ is the system's self-energy. For the free-fermion bath,
the retarded component of $\Sigma_b$ is explicitly given in the frequency domain as
\begin{align}
\Sigma_b^R(\nu)
&=
\sum_k \frac{V_k^2}{\nu-\varepsilon_k+i\eta},
\end{align}
where $\eta$ is an infinitesimal positive constant. By using $1/(\nu+i\eta)=\mathcal P(1/\nu)-i\pi\delta(\nu)$
(with $\mathcal P$ being the principal value), one obtains
the imaginary part of $\Sigma_b^R(\nu)$,
\begin{align}
\Gamma(\nu)
&=
\sum_k \pi V_k^2 \delta(\nu-\varepsilon_k),
\end{align}
which corresponds to the bath's spectral function. One often assumes a flat density of states for the bath, $\Gamma(\nu)=\Gamma$,
as the simplest choice. The real part of $\Sigma_b^R(\nu)$ gives a potential shift, which can be renormalized into the chemical potential.
The advanced component is simply the complex conjugate, $\Sigma_b^A(\nu)=\Sigma_b^R(\nu)^\dagger$.
The Keldysh part of $\Sigma_b$ is determined from the fluctuation-dissipation relation for the bath's Green's function,
giving $\Sigma_b^K(\nu)=F(\nu)[\Sigma_b^R(\nu)-\Sigma_b^A(\nu)]$ with $F(\nu)=\tanh(\beta\nu/2)$
($\beta$ is the inverse temperature of the heat bath). 

To summarize, the self-energy due to the coupling to the free-fermion bath takes a simple form,
\begin{align}
\begin{pmatrix}
\Sigma_b^R(\nu) & \Sigma_b^K(\nu) \\
0 & \Sigma_b^A(\nu)
\end{pmatrix}
&=
\begin{pmatrix}
-i\Gamma & -2i\Gamma F(\nu) \\
0 & i\Gamma
\end{pmatrix}.
\end{align}
Although the free-fermion bath model looks somewhat artificial, it correctly reproduces the semiclassical result of Boltzmann's transport theory
\cite{Han2013, Aoki2014}.
In this context, the model can be viewed as a two-parameter phenomenology (the relaxation rate $\Gamma$ and heat bath's temperature $\beta^{-1}$) 
to describe dissipation. 
One can also employ other models for a heat bath such as a model of phonons, but then 
the bath's degrees of freedom cannot be integrated out
analytically, and one has to use certain approximations.

One can incorporate the effect of periodic driving in this approach. To this end, one assumes the existence of time-periodic nonequilibrium steady states.
Once the steady state is reached, the Green's function should satisfy the periodicity in two-time arguments, $G^\alpha(t,t')=G^\alpha(t+T,t'+T)$
($\alpha=R,A,K$).
This allows one to define the Floquet Green's function \cite{Faisal1989, Martinez2003, Martinez2005, TsujiOkaAoki2008, Aoki2014, Liu2017},
\begin{align}
&
G_{mn}^\alpha(\nu)
\notag
\\
&:=
\frac{1}{T}\int_0^T ds \int_{-\infty}^\infty dt\, e^{i(m-n)\omega s+i\nu t}
G^\alpha\left(s+\frac{t}{2}, s-\frac{t}{2}\right),
\end{align}
which takes a matrix form with indices $m$ and $n$ being the labels of Fourier modes as used in Eq.~(\ref{H_mn}).
The Dyson equation maintains the form of Eq.~(\ref{Dyson equation}) with all the Green's functions and self-energies 
written in the above matrix form with the inverse interpreted as the matrix inverse. The Floquet matrix form of $\Sigma_b$ reads
$\Sigma_{b,mn}^R(\nu)=-i\Gamma\delta_{mn}$ and $\Sigma_{b,mn}^K(\nu)=-2i\Gamma F(\beta(\nu+n\omega)/2)\delta_{mn}$.

\begin{figure}[t]
\includegraphics[width=8.5cm]{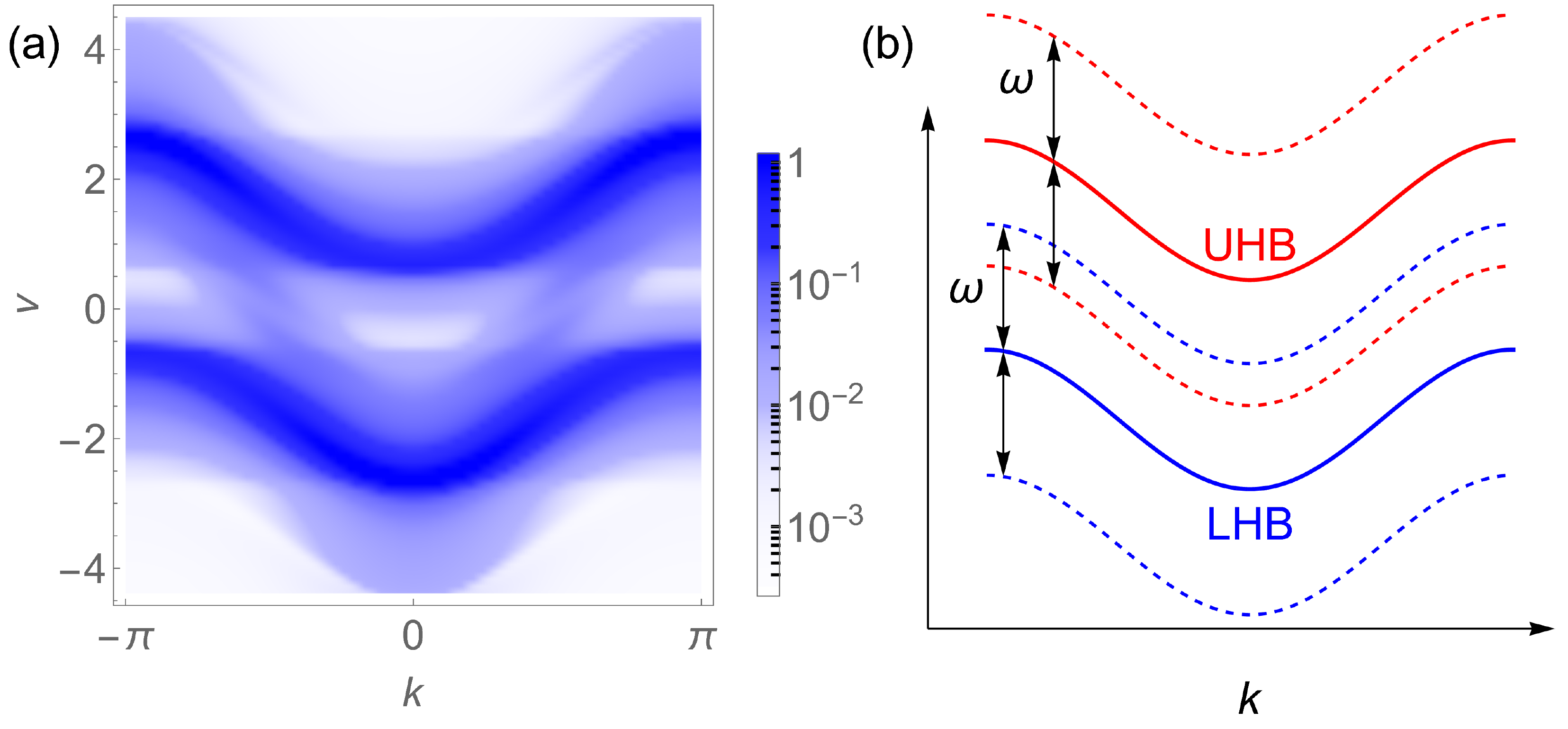}
\caption{(a) The single-particle spectrum $A(\nu,k)$ of the ac-field-driven Falicov-Kimball model calculated from the Floquet DMFT.
(b) The corresponding schematic band structure of the ac-field-driven Mott insulator with the drive frequency $\omega$. 
The upper and lower Hubbard bands (UHB and LHB) are shown by solid curves, and the accompanied Floquet sidebands
are shown by dashed curves. (Adapted from Aoki et al. (2014) Nonequilibrium dynamical mean-field theory and its applications.
{\it Rev. Mod. Phys.} 86: 779 \cite{Aoki2014}.)}
\label{FK spectrum}
\end{figure}

Provided that the self-energy from the system's interaction is given, 
the Green's function in the time-periodic nonequilibrium steady state is determined
from the Dyson equation (\ref{Dyson equation}).
This part requires a (diagrammatic) solver for quantum many-body problems.
One such approach is the so-called Floquet dynamical mean-field theory (DMFT) 
\cite{TsujiOkaAoki2008, TsujiOkaAoki2009, JouraFreericksPruschke2008, LubatschKroha2009, Aoki2014},
which is a combined formulation of the Floquet theory
and nonequilibrium DMFT \cite{FreericksTurkowskiZlatic2006, Aoki2014}. 
The latter solves nonequilibrium quantum many-body problems under the local approximation of the self-energy
\cite{GeorgesKotliarKrauthRozenberg1996}.

In Fig.~\ref{FK spectrum}, an example of the result of the Floquet DMFT is shown for the ac-field-driven Falicov-Kimball model
\cite{TsujiOkaAoki2008, TsujiOkaAoki2009, Aoki2014}, 
which is a simple model of correlated electrons that shows a Mott insulator-to-metal transition in equilibrium \cite{FreericksZlatic2003}.
There emerges an ac-field-induced midgap state between the upper and lower Hubbard bands
as Floquet sidebands with an energy spacing $\omega$. Due to this, the system turns from a Mott-insulating state
to an ac-field-induced metallic state. This exemplifies a quantum many-body Floquet state.

The nonequilibrium Green's function approach not only describes the single-particle properties, but also
characterizes two-particle ones such as response functions and transport coefficients.
To obtain those quantities, one has to evaluate two-particle Green's functions on top of the single-particle Green's function.
For example, the transport property of the ac-field-driven Falicov-Kimball model can be identified from the optical conductivity spectrum,
which shows a Drude-like peak structure at low energy instead of a Mott gap, indicating that the system indeed becomes metallic
due to the effect of the ac field \cite{TsujiOkaAoki2009}.

\section{Experimental observations}

\begin{figure}[t]
\includegraphics[width=8cm]{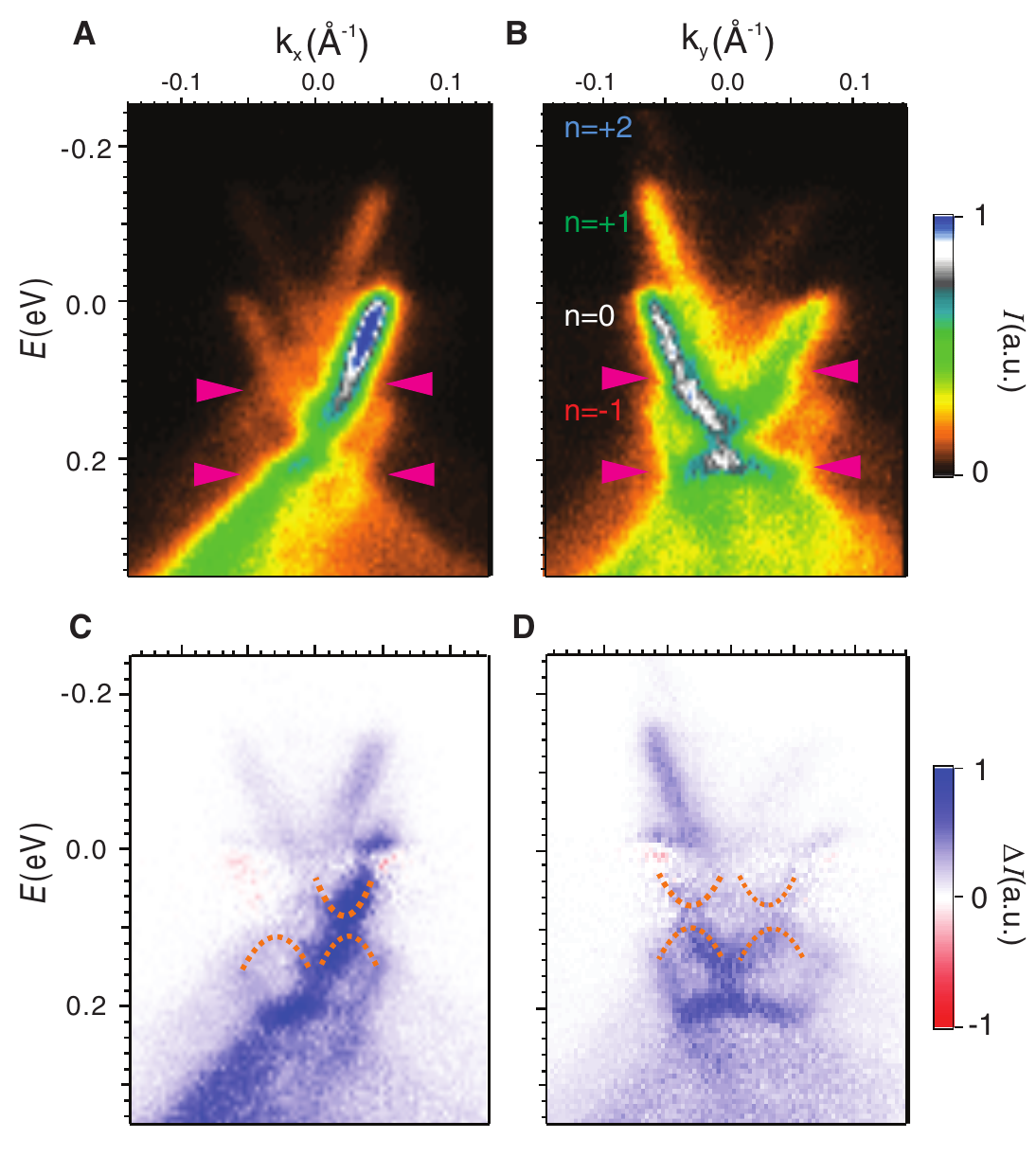}
\caption{Time-resolved ARPES spectra of surface states of a topological insulator Bi$_2$Se$_3$ under circularly polarized laser excitation
along (A) $k_x$ and (B) $k_y$ directions. (C) and (D) are the corresponding spectra after subtracting the spectra at $t=-500$ fs.
(Adapted from Wang et al. (2013) Observation of Floquet-Bloch States on the Surface
of a Topological Insulator. {\it Science} 342: 453-457 \cite{Wang2013}.)}
\label{arpes}
\end{figure}

Floquet states have been experimentally observed in various systems.
In solid-state materials, occupied single-particle spectra can be measured by angle-resolved photoemission spectroscopy (ARPES),
which can map out the band structure of correlated electrons in energy and momentum spaces below the Fermi level.
The technique can also be applied to study nonequilibrium states of electrons in solids, in which case pump and probe pulses
are applied with a certain time delay in between (time-resolved ARPES). 
The former pulse is used to drive electrons out of equilibrium, while the latter is 
used to generate photoelectrons. During irradiation of a pump pulse, the system can be approximately considered to be driven
periodically in time if the pump pulse contains sufficiently large number of oscillation cycles.
Thus, the time-resolved ARPES offers an opportunity to observe Floquet states for electron systems.

In Fig.~\ref{arpes}, an example of time-resolved ARPES spectra is shown for a topological insulator Bi$_2$Se$_3$ driven
by a circularly polarized laser \cite{Wang2013}. 
In topological insulators, there are topologically protected gapless surface states,
which support an energy dispersion of Dirac cones. This is precisely the same situation as we see in Sec.~\ref{sec: examples}:
Dirac electrons driven by circularly polarized light undergoes a transition to a Floquet topological insulator with a gap opening
at the Dirac point. In the time-resolved ARPES experiment, the energy gap is observed to appear
under the excitation of a circularly polarized laser. On top of that, Floquet sideband structures are observed 
with the energy spacing of the drive frequency in the ARPES spectra. These are consistent with the results of the Floquet theory
\cite{OkaAoki2009}
(see, e.g., Fig.~\ref{dirac}).
The subsequent experiment has also demonstrated the transition from the Floquet states to Volkov states,
i.e., photon-dressed states of free electrons, by time-resolved ARPES \cite{Mahmood2016}.

When Dirac electrons are driven by circularly polarized light and turn to a Floquet topological insulator, the bands
around the Dirac point can acquire non-zero Chern numbers.
As a result, the system shows the anomalous Hall effect, that is, current flows perpendicular to the direction of an applied dc electric field
without a magnetic field.
If the distribution of the occupied states are not far from that of zero-temperature equilibrium states,
the Hall conductance approaches the quantized value. The Hall conductance of Dirac electrons in a monolayer graphene induced by
a circularly polarized laser has been measured experimentally, using an on-chip photoconductive switch device
\cite{McIver2020}.
The result shows that Floquet topological bands indeed generate the anomalous Hall effect, 
demonstrating characteristic features of Floquet topological states.

\begin{figure}[t]
\includegraphics[width=7.5cm]{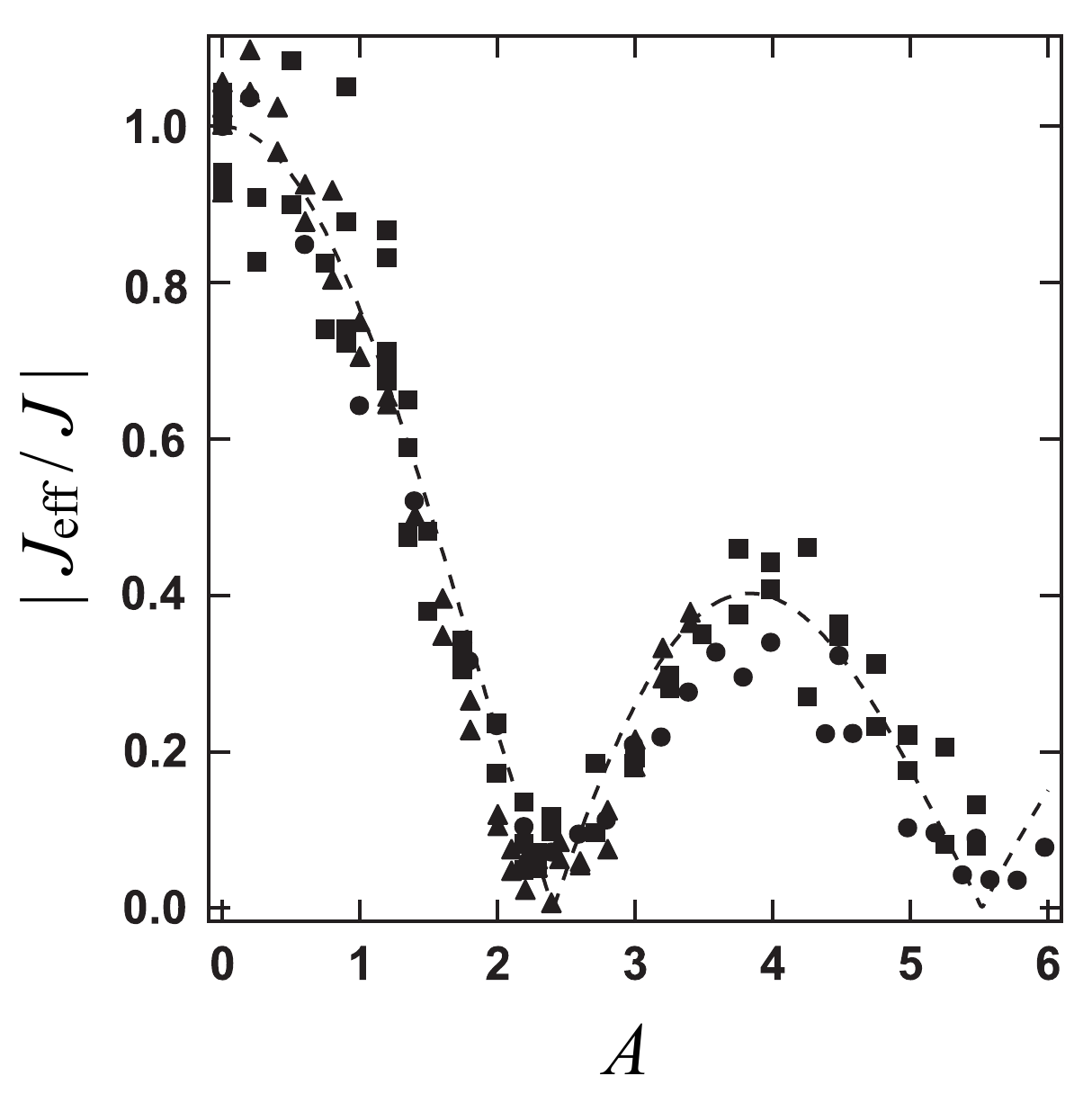}
\caption{Effective hopping amplitude of Bose-Einstein condensed cold atoms
in a periodically shaken optical lattice as a function of the amplitude of the periodic drive. 
The dashed curve represents the Bessel function $\mathcal J_0(A)$.
(Adapted from Lignier et al. (2007) Dynamical Control of Matter-Wave Tunneling in Periodic Potentials.
{\it Phys. Rev. Lett.} 99: 220403 \cite{Lignier2007}.)}
\label{bec}
\end{figure}

In contrast to solid-state systems, artificial quantum systems provide high controllability and ideal clean situations
to realize Floquet states. For example, cold atomic gases trapped in an optical lattice are well described by
a lattice model of isolated quantum many-body systems.
One can then drive the system by periodically shaking the optical lattice potential in real space \cite{Eckardt2017}, 
mimicking the effect of ac electric fields for electrons.

In Fig.~\ref{bec}, the experimentally measured effective hopping amplitude of Bose-Einstein condensates 
in a periodically shaken optical lattice is shown as a function of the amplitude $A$ of the periodic drive \cite{Lignier2007}.
As we see in Sec.~\ref{sec: examples} and \ref{sec: high-frequency expansion}, 
the Floquet theory predicts that the hopping amplitude is renormalized by
the Bessel factor as $J_{\rm eff}=J\mathcal J_0(A)$, which perfectly agrees with the experimental observation.
Using this effect, one can effectively control the interaction strength with ac fields, inducing, for instance, 
a superfluid-to-Mott insulator transition in the Bose-Hubbard model \cite{Eckardt2005}.
One can even reverse the sign of the hopping if $\mathcal J_0(A)<0$,
which can be used to realize frustrated classical spin systems \cite{Struck2011} or to effectively convert the interaction from repulsive to attractive
in the ac-field-driven Fermi-Hubbard model \cite{TsujiOkaWernerAoki2011}.

Periodic shaking of optical lattices has also been used to realize the Haldane model in fermionic cold-atom systems \cite{Jotzu2014},
where periodic circular shaking 
induces the Floquet topological insulator with complex hopping parameters on a honeycomb lattice as discussed above.
The generated band gap is probed by momentum-resolved measurement of interband transitions.
The effect of Berry curvature has been demonstrated by applying a constant force to atoms and finding an orthogonal drift,
in the similar manner as the anomalous Hall effect.
The induced Berry curvature has been reconstructed in the entire Brillouin zone experimentally \cite{Flaschner2016}.

Other physical systems that allow experimental observations of Floquet states include
photonic systems \cite{Kitagawa2012, Rechtsman2013, Maczewsky2017, Mukherjee2017, Ozawa2019}, 
trapped ions \cite{Kiefer2019}, and superconducting qubits \cite{Deng2015, Roushan2017}.

\section{Summary and outlook}

Here we have reviewed basic aspects of Floquet states and their applications to various periodically driven quantum systems.
It allows one to control phases of matter dynamically (`Floquet engineering' \cite{Bukov2015, OkaKitamura2019, Weitenberg2021}) 
that may not be realized
in equilibrium.
While our understanding of Floquet states has been deepened, 
the inherent potential of Floquet states has not been fully elaborated experimentally, especially in solid-state systems.
This may be partly because (i) the required amplitude of electric fields is often large,
(ii) one needs ultrafast time resolution to capture driven states during irradiation of a pump pulse,
(iii) transitions to higher bands and modulation of the distribution may occur in real materials, 
(iv) heating will suppress quantum fluctuations, and
(v) energy dissipation takes place in complicated ways.

While these will make it a great challenge to realize Floquet states in solids,
they can also offer opportunities: 
The development of new materials and new devices will pave a way to explore the frontier of Floquet states.
For example, a recent experiment has shown an ideal behavior of driven two-level systems with the scaling of the Bessel factor
up to extremely high orders in a diamond nitrogen-vacancy center \cite{Nishimura2022}. 
If the interaction between the two-level systems can be implemented in those systems,
it will provide an interesting playground of Floquet many-body states. 
Another opportunity is the effect of dissipation in open quantum systems, which can work in a favorable direction 
to realize (rather than suppress) new orders and new states of matter in quantum systems.
For example, a two-body loss due to inelastic collisions in the Fermi-Hubbard model can effectively change the antiferromagnetic interaction
to the ferromagnetic one \cite{Nakagawa2020}, which has been recently observed in cold-atom systems \cite{Honda2023}.
Finally, there has been an experimental progress in probing highly oscillating electronic states 
through high harmonic generation in solids \cite{Ghimire2011, Schubert2014, Luu2015, Vampa2015, Hohenleutner2015},
which will be a useful key tool to study Floquet states in the future.










\bibliography{ref}

\end{document}